\documentclass[aps,pra,twocolumn,showpacs,superscriptaddress,longbibliography]{revtex4-2}
\usepackage{graphicx} %Include figure files,superscriptaddress
\usepackage{amsmath}
\usepackage{multirow}
\usepackage{graphicx,epstopdf}
\usepackage{booktabs}
\usepackage{gensymb}
\usepackage[dvipsnames]{xcolor}
\usepackage{footnote}
\usepackage{multibib}

\newcommand{\be}{\begin{equation}}
	\newcommand{\ee}{\end{equation}}
\newcommand{\bea}{\begin{eqnarray}}
	\newcommand{\eea}{\end{eqnarray}}
\newcommand{\bse}{\begin{subequations}}
	\newcommand{\ese}{\end{subequations}}

\graphicspath{{../}}

\usepackage{color}
\usepackage[colorlinks,bookmarks=false,citecolor=darkblue,linkcolor=red,urlcolor=blue]{hyperref}

\definecolor{darkred}{rgb}{0.7,0.0,0.0}

\definecolor{darkblue}{rgb}{0,0.02,0.45}

\definecolor{darkgreen}{rgb}{0.02,0.45,0.0}

\definecolor{violet}{rgb}{0.8,0.2,0.6}

\begin{document}
\title{Collinear antiferromagnetic order in spin-$\frac52$ triangle lattice antiferromagnet Na$_3$Fe(PO$_4$)$_2$}

\author{Sebin J. Sebastian}
\affiliation{School of Physics, Indian Institute of Science Education and Research Thiruvananthapuram-695551, India}
\author{A. Jain}
\author{S. M. Yusuf}
\affiliation{Solid State Physics Division, Bhabha Atomic Research Centre, Mumbai 400 085, India}
\author{M. Uhlarz}
%\author{Y. Skourski}
\affiliation{Dresden High Magnetic Field Laboratory (HLD-EMFL), Helmholtz-Zentrum Dresden-Rossendorf, 01328 Dresden, Germany}
\author{R. Nath}
\email{rnath@iisertvm.ac.in}
\affiliation{School of Physics, Indian Institute of Science Education and Research Thiruvananthapuram-695551, India}
\date{\today}
\begin{abstract}
We set forth the structural and magnetic properties of the frustrated spin-$5/2$ triangle lattice antiferromagnet Na$_3$Fe(PO$_4$)$_2$ examined via x-ray diffraction, magnetization, heat capacity, and neutron diffraction  measurements on the polycrystalline sample. No structural distortion was detected from the temperature-dependant x-ray diffraction down to 12.5~K, except a systematic lattice contraction. The magnetic susceptibility at high temperatures agrees well with the high-temperature series expansion for a spin-$5/2$ isotropic triangular lattice antiferromagnet with an average exchange coupling of $J/k_{\rm B} \simeq 1.8$~K rather than a one-dimensional spin-$5/2$ chain model. This value of the exchange coupling is consistently reproduced by the saturation field of the pulse field magnetization data. It undergoes a magnetic long-range-order at $T_{\rm N} \simeq 10.4$~K. Neutron diffraction experiments elucidate a collinear antiferromagnetic ordering below $T_{\rm N}$ with the propagation vector $k = (1,0,0)$. An intermediate value of frustration ratio ($f \simeq 3.6$) reflects moderate frustration in the compound which is corroborated by a reduced ordered magnetic moment of $\sim 1.52$~$\mu_{\rm B}$ at 1.6~K, compared to its classical value ($5~\mu_{\rm B}$). Magnetic isotherms exhibit a change of slope envisaging a field induced spin-flop transition at $H_{\rm SF}\simeq3.2$~T. The magnetic field vs temperature phase diagram clearly unfold three distinct phase regimes, reminiscent of a frustrated magnet with in-plane (XY-type) anisotropy.
\end{abstract}
%\pacs{75.50.Ee, 71.20.Ps, 75.10.Pq, 75.30.Kz, 75.30.Et, 75.10.Jm}
\maketitle

\section{Introduction}
Triangular-lattice antiferromagnet (TLA) is a simplest example of geometrically frustrated magnets which has been a motif since the advent of various exotic quantum phases including quantum spin-liquid (QSL)~\cite{Collins605,*Starykh052502,Balents199}. Ideally, two-dimensional (2D) TLAs with isotropic (Heisenberg) exchange interaction undergo a non-collinear $120^{\circ}$ ordering irrespective of their spin value~\cite{Jolicoeur2727,Chubukov8891,Capriotti3899,Bernu10048,Chernyshev144416}. However, in real materials the inherent exchange anisotropy and inter-layer couplings often have broad implications, leading to more complex and non-trivial ground states~\cite{Kawamura4138}. For instance, TLAs with easy-axis anisotropy, the low temperature 120$^{\circ}$ state is often preceded by a collinear state in zero-field, as in Ba$_3$(Mn,Co)Nb$_2$O$_9$ and other TLA compounds~\cite{Miyashita3385,Melchy064411,Clark788,Kadowaki751, Ajiro4142,Harrison679,Ranjith115804,Lee224402,*Lee104420} while for systems with easy-plane anisotropy, the collinear phase is unstable and entail only $120^{\circ}$ ordering~\cite{Miyashita3605,Rawl054412}. The materials with noncollinear magnetic structure give rise to spin chirality, a wellspring of strong magnetoelectric coupling~\cite{Cheong13}. Moreover, spatially anisotropic spin-$1/2$ TLA model has been proposed to host QSL phase for a critical value of the ratio of exchange couplings~\cite{Zhu157201,Zhu207203,Heidarian012404}. Recently, an extended fluctuating regime with slow dynamics below magnetic ordering is detected in the spin-$3/2$ isotropic TLAs~\cite{Somesh104422}. Overall, diverse phase regimes are anticipated in TLAs while going from a classical (high spin) to quantum (low spin) limit owing to the geometrical frustration in concurrence with magnetic anisotropy, effect of fluctuations etc~\cite{Yamamoto4263,Ranjith014415}.

\begin{figure}
\includegraphics[width=\columnwidth]{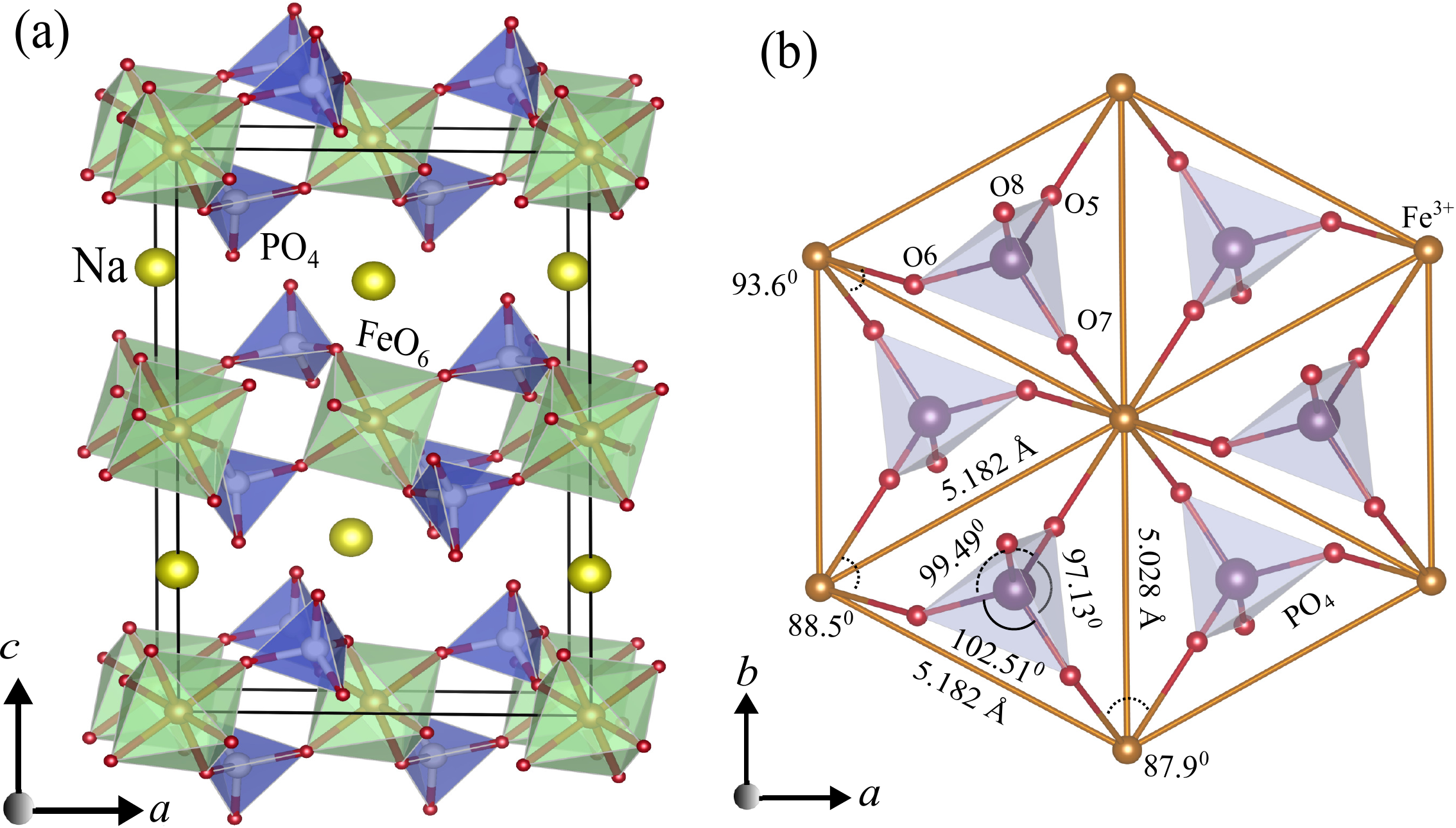}
\caption{\label{Fig1}(a) Crystal structure of Na$_3$Fe(PO$_4)_2$ showing triangular layers sculpted by the corner-shared FeO$_6$ octahedra and PO$_4$ tetrahedra. Location of Na$^{+}$ ions is also shown. (b) A section of the layer showing Fe$^{3+}$ ions linked via PO$_4$ tetrahedra on an anisotropic triangular lattice.
%(c) A schematic representation of magnetic exchange couplings ($J$ and $J'$) on a triangular layer formed by Fe$^{3+}$ ions.
}
\end{figure}
Recently, a series of compounds with general formula $AA^{\prime}M$($X$O$_4$)$_2$ ($A$~=~Ba, K, Rb, and Sr, $A^{\prime}=$ Na$_2$ and Ag$_2$,  $M=$ Mn, Ni, Co, Cr, Fe, and Cu, and $X$ = P and V) are reported with varying crystal structures~\cite{Sebastian064413, Amuneke2207,Amuneke5930,Moller214422,Nakayama116003,Reub6300,Sanjeewa2813,Lee224420,Tsirlin014401}. Interestingly, this series of compounds are projected as low-dimensional spin systems where the superexchange takes place in an elongated path involving VO$_4$ or PO$_4$ units. Though majority of the compounds in this series are described in terms of anisotropic triangular lattice model, yet some of them are found to be exceptions. BaAg$_2$Cu(VO$_4$)$_2$ with triclinic ($P\bar 1$) structure was initially proposed to be a spin-$1/2$ anisotropic triangular lattice system, due to its unequal Cu--Cu bond lengths~\cite{Amuneke2207}. Subsequently, the band structure calculations ruled out this proposal and established a superposition of double spin-$1/2$ chains: one is ferromagnetic and another is antiferromagnetic~\cite{Tsirlin014401}. Similarly, the monoclinic ($C2/c$) compound BaNa$_2$Cu(VO$_4$)$_2$ features antiferromagnetic spin-$1/2$ crossed chains arranged 62${^\circ}$ with respect to each other~\cite{Sebastian064413}.
%Similarly, a chromium-based ATL system, Ag$_3$Cr[VO$_4$]$_2$ have demonstrated a significant suppression of LRO till 10~K and was pushed to the least temperature 0.031~K by tuning the spacer atoms ($A$= K, Rb)~\cite{Tapp064404}.
Moreover, the signature of QSL is reported in the structurally perfect spin-$1/2$ TLA BaNa$_2$Co(PO$_4)_2$~\cite{Lee024413} and spin-$3/2$ TLAs (K,Rb)Ag$_2$Cr(VO$_4$)$_2$~\cite{Tapp064404,Lee224420}, piquing further interest in this series of compounds. Thus, this series of materials serve as a cradle for novel ground states where one can tune the ground state properties by changing the chemical pressure.

In the following, we delineate a comprehensive study of the thermodynamic properties of the frustrated spin-$5/2$ magnet Na$_3$Fe(PO$_4)_2$ in which the spins are embedded in a triangular lattice. It crystallizes in a monoclinic space group ($C2/c$) and belongs to the family $AA^{\prime}M$($X$O$_4$)$_2$ ~\cite{Morozov377,*Belkhiria117}. The crystal structure is presented in Fig.~\ref{Fig1} where the slightly distorted FeO$_6$ octahedra are corner shared with distorted PO$_4$ tetrahedra in order to form the triangular Fe$^{3+}$ ($3d^5$, spin-$5/2$) layers in the $ab$-plane which are stack along the $c$-direction. Isolated Na atoms are found between the layers. In each layer, the Fe$^{3+}$  triangles are slightly anisotropic with two Fe$^{3+}$-Fe$^{3+}$ distances of 5.182\,\r A and one of 5.028\,\r A~[Fig.~\ref{Fig1}(b)]. Further considering the analogy that the shortest Fe$^{3+}$-Fe$^{3+}$ distance leads to a strong exchange interaction and the longer distance leads to a weak interaction, the spin lattice can also be viewed as coupled spin chains. Of course, the magnitude and sign of the interactions may vary depending on the actual path involved.
%Figure~\ref{Fig1}(c) illustrates a section of the spin-lattice where chains with intrachain coupling $J$ along $b$-direction are coupled by $J'$ in the $ab$-plane.
Our magnetization data could be described well using the spin-$5/2$ isotropic triangle lattice model. It shows the onset of a magnetic ordering at $T_{\rm N} \simeq 10.4$~K which is incisively detected to be a collinear AFM type. The $H-T$ phase diagram with a spin-flop (SF) transition confirms magnetic anisotropy in the compound.

\section{Methods}
Polycrystalline sample of Na$_3$Fe(PO$_4)_2$ was synthesized following the conventional solid-state reaction technique. Prior to synthesis, the precursor FePO$_4$ was prepared by heating stoichiometric mixture of Fe$_2$O$_3$ (Aldrich, 99.99\%) and (NH$_4$)$_2$HPO$_{4}$ (Aldrich, 99.995\%) at 880~$^{0}$C for 12 hours in air. In the next step, the raw materials Na$_{2}$CO$_{3}$ (Aldrich, 99.995\%) and (NH$_4$)$_2$HPO$_{4}$ (Aldrich, 99.995\%) were mixed with FePO$_4$ in stoichiometric ratios, grounded thoroughly, and pressed into pellets. The pellets were then sintered in an alumina crucible at 730~$^{0}$C for three days in air with several intermediate grindings.

The phase purity of the sample was confirmed from the powder x-ray diffraction (XRD) performed at room temperature. For the powder XRD experiment, a PANalytical powder diffractometer with Cu\textit{K}$_{\alpha}$ radiation ($\lambda_{\rm avg} \simeq 1.54182$~{\AA}) was used. The temperature-dependent powder XRD measurement was performed over the temperature range 12.5~K~$\leq T \leq 300$~K using a low-temperature attachment (Oxford Phenix) to the diffractometer. The dc magnetization ($M$) was measured on the powder sample using a superconducting quantum interference device (SQUID) (MPMS-3, Quantum Design) magnetometer. The measurements were performed in the temperature range 1.8~K~$\leq T \leq 380$~K and in the magnetic field range 0~$\leq H \leq 7$~T. High-field magnetization was measured in pulsed magnetic field at the Dresden high magnetic field laboratory. The details of the experimental procedure are described in Ref.~\cite{Skourski214420,*Tsirlin132407}. Heat capacity as a function of temperature and magnetic field was measured on a small pellet using a Physical Property Measurement System (PPMS, Quantum Design).

Neutron powder diffraction (NPD) experiments were carried out in the temperature range 1.6~K~$\leq T \leq 15$~K using the powder diffractometer (PD-II, $\lambda \simeq 1.2443$~{\AA}) at Dhruva reactor, Bhabha Atomic Research Centre (BARC), Mumbai, India. The 1D neutron-depolarization measurements were performed using the polarized neutron spectrometer (PNS) at the Dhruva reactor, BARC, Mumbai ($\lambda \simeq 1.205$~{\AA}). For this experiment, polarized neutron beam was produced and analysed using the magnetized Cu$_2$MnAl [reflection from $(111)$ plane] and Co$_{0.92}$Fe$_{0.8}$ [reflection from $(200)$ plane] single crystals, respectively. The two states of the polarization of incident neutron beam were achieved by a flipper placed just before the sample. The flipping ratio ($R$) of the neutron beam was determined by measuring the intensities of neutrons in non-spin flip and spin flip channels with the flipper on and off, respectively.
Rietveld refinement of the powder XRD and NPD data was performed using \verb"FullProf" software package~\cite{Rodriguez55}.
%Inorder to obtain the magnetic structure, temperature-dependant neutron powder diffraction (NPD) was done on the powder sample in Bhabha Atomic Research Center (BARC), India. The experiment was done using the neutron powder diffractometer PD-I ($\lambda \simeq 1.2443$~{\AA}) with three linear position-sensitive detectors at Dhruva reactor.

Quantum Monte-Carlo (QMC) simulation and full diagonalization were performed using the \verb|loop| and \verb|fulldiag| algorithms~\cite{Todo047203} of the ALPS package~\cite{Albuquerque1187}. 

\section{Results}
\subsection{X-ray Diffraction}
\begin{figure}
\centering
\includegraphics[width=\columnwidth]{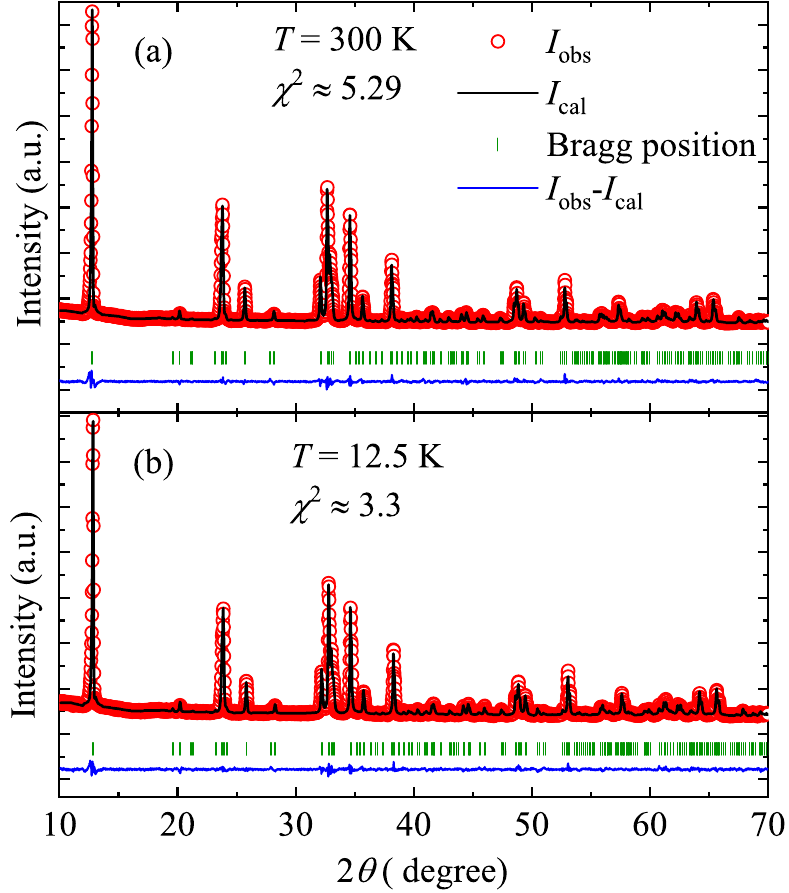}
\caption{\label{Fig2} Powder XRD pattern of Na$_3$Fe(PO$_4)_2$ measured at (a) $T = 300$~K and (b) $T = 12.5$~K. The open circles are the experimental data and the solid black line is the Rietveld refined fit. The Bragg positions are indicated by green vertical bars and the bottom solid blue line indicates the difference between the experimental and calculated intensities.}
\end{figure}
\begin{figure}
\centering
\includegraphics[width=\columnwidth]{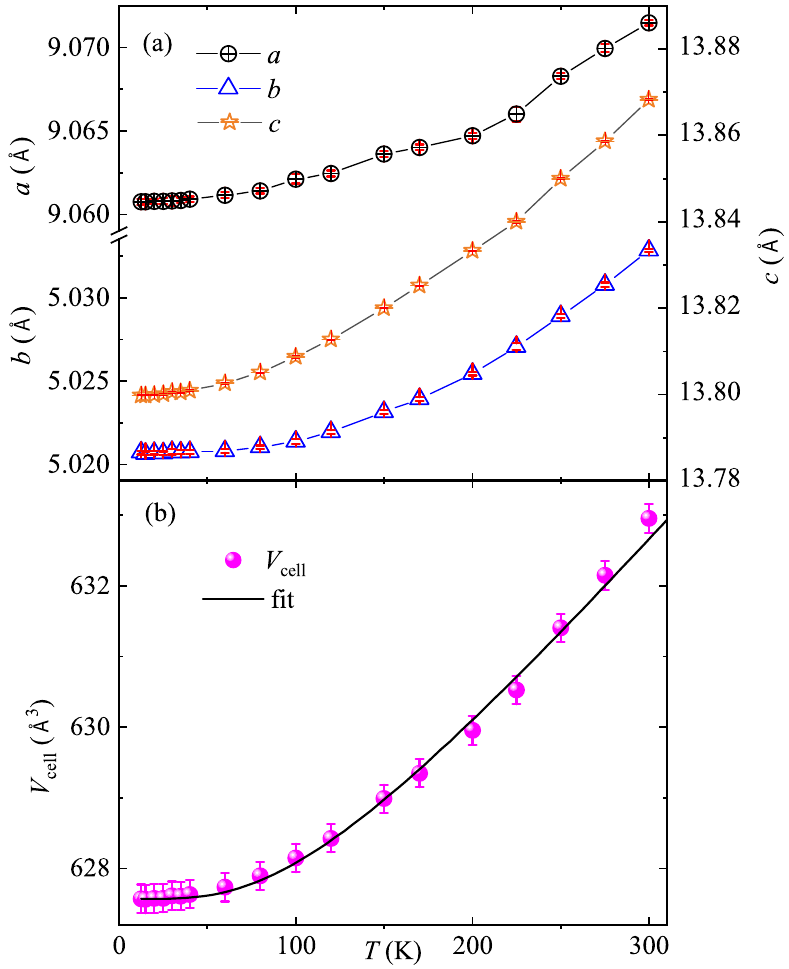}
\caption{\label{Fig3} The lattice constants ($a$, $b$, and $c$), and unit cell volume ($V_{\rm cell}$) are plotted as a function of temperature from 12.5 to 300~K. The solid line in (b) represents the fit using Eq.~\eqref{eq1}.}
\end{figure}
In order to check the structural phase transition or lattice distortion, the sample was examined by measuring XRD pattern at different temperatures.
Figure~\ref{Fig2} displays the powder XRD pattern at two end temperatures along with the Rietveld fits. The obtained lattice parameters from the refinement at room temperature are $a = 9.0714(2)$~{\AA}, $b = 5.032(1)$~{\AA}, $c = 13.8683(3)$~{\AA}, and $\beta = 91.44(1)^{\circ}$ and the unit cell volume $V_{\rm cell}\simeq 632.96$~{\AA}$^{3}$, which are in close agreement with the previous report~\cite{Morozov377}. Temperature variation of the lattice constants ($a$, $b$, and $c$) and $V_{\rm cell}$ are shown in Fig.~\ref{Fig3}. All of them are found to decrease systematically upon cooling down to 12.5~K, as expected. No anomaly at the temperature corresponding to the magnetic transition suggests the absence of significant magnetoelastic coupling.

Such a temperature dependence of $V_{\rm cell}$ is usually described well by~\cite{Islam174432}
\begin{equation}
V(T)=\gamma U(T)/K_0+V_0,
\label{eq1}
\end{equation}
where $V_0$ is the cell volume in the zero temperature limit, $K_0$ is the bulk modulus, and $\gamma$ is the Gr$\ddot{\rm u}$neisen parameter. $U(T)$ is the internal energy, which can be derived in terms of the Debye approximation as
\begin{equation}
U(T)=9nk_{\rm B}T\left(\frac{T}{\theta_{\rm D}}\right)^3\int_{0}^{\theta_{\rm D}/T}\dfrac{x^3}{e^x-1}dx.
\label{eq2}
\end{equation}
Here, $n$ is the number of atoms in the unit cell and $k_{\rm B}$ is the Boltzmann constant. As shown in Fig.~\ref{Fig3}(b), $V_{\rm cell}(T)$ above 12.5~K was fitted reasonably by Eq.~\eqref{eq1} producing the parameters: Debye temperature $\theta_{\rm D} \simeq 400$~K, $\gamma/K_0 \simeq 8.41 \times 10^{-5}$~Pa$^{-1}$, and $V_0 \simeq 627.6$~\AA$^{3}$.

\subsection{Magnetization}
\begin{figure}
\includegraphics[width=\columnwidth]{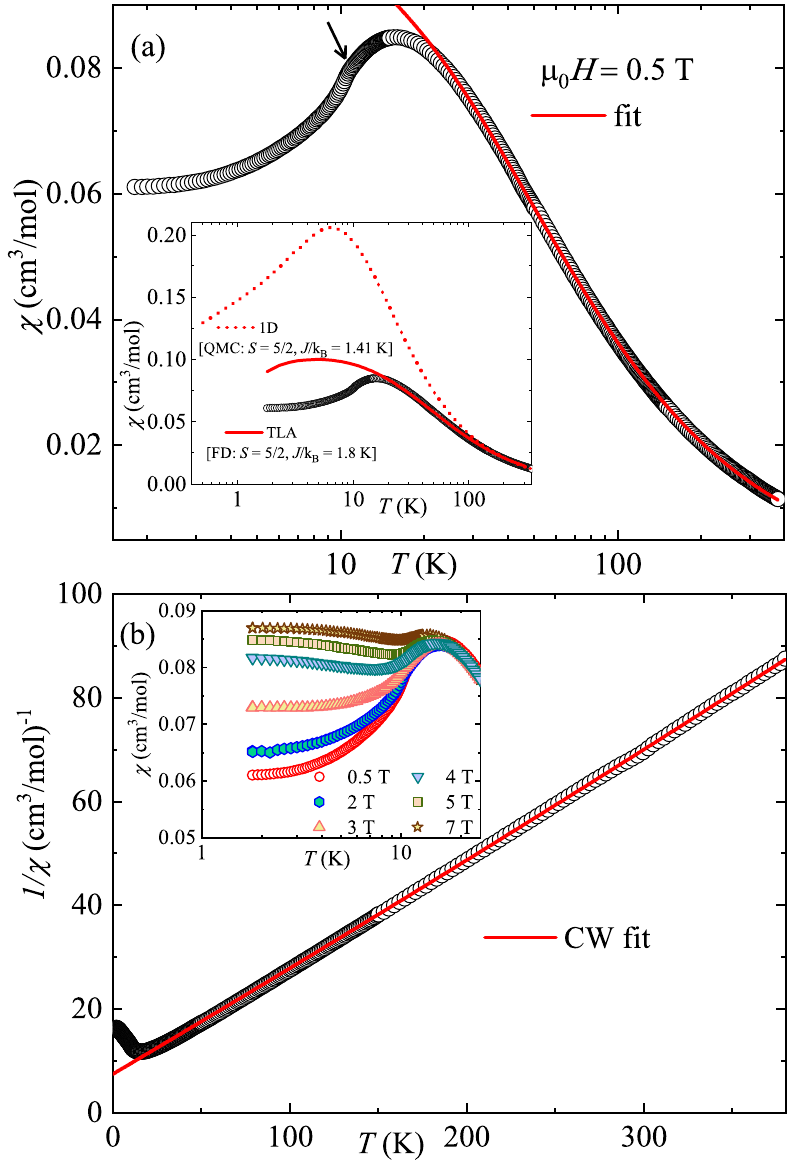}
\caption{\label{Fig4} (a) $\chi$ as a function of temperature in an applied field $\mu_0H = 0.5$~T. The downward arrow points to the transition temperature ($T_{\rm N}$). The solid line represents the fit using isotropic triangular lattice model [Eq.~\eqref{eq4}]. Inset: comparison of experimental data with spin-$5/2$ uniform chain model and isotropic triangular lattice simulated using QMC and full diagonalization (FD), respectively. (b) Inverse susceptibility $1/\chi$ vs $T$ and the solid line represents the CW fit, as discussed in the text. Inset: $\chi(T)$ measured in different fields, below $T_{\rm N}$.}
\end{figure}
\begin{figure}
\includegraphics[width=\columnwidth] {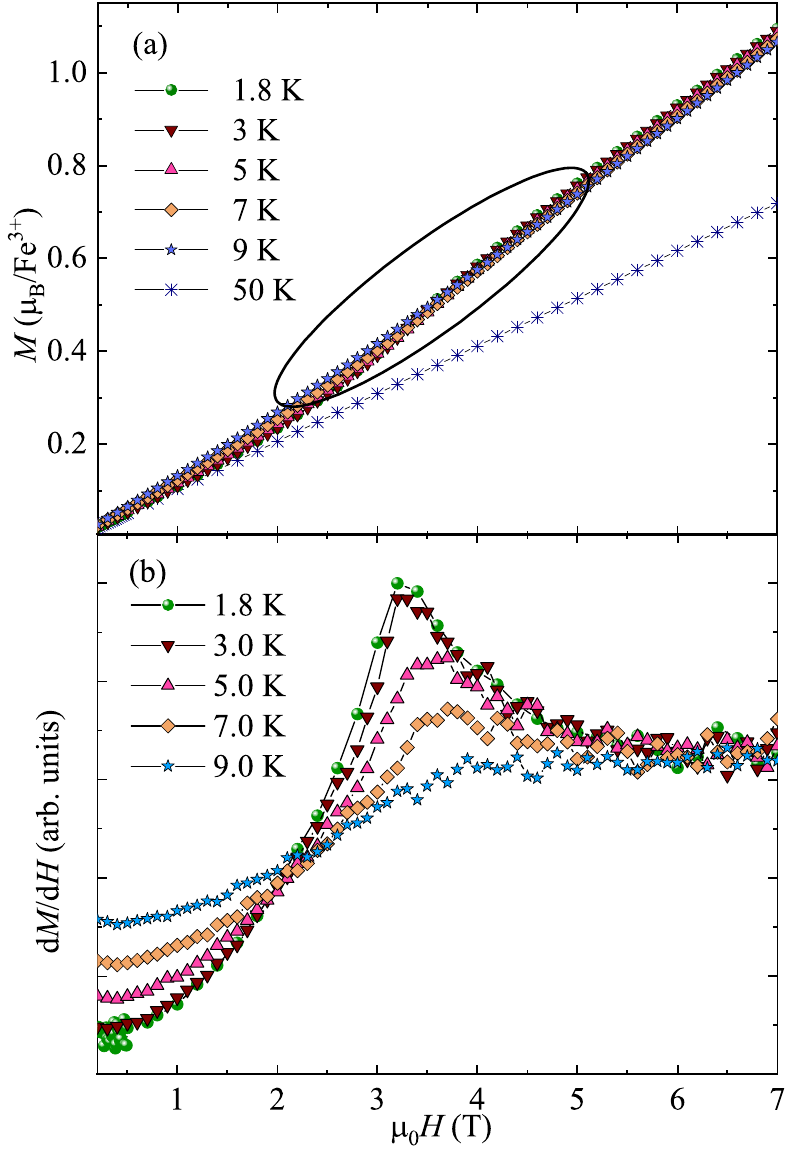}
\caption{\label{Fig5} (a) Magnetization ($M$) vs $H$ measured at various temperatures from 1.8~K to 50~K. The encircled portion depicts the field induced transition regime. (b) The derivative $dM/dH$ vs $H$ is plotted at various temperatures from 1.8~K to 9~K in-order to highlight the field induced transition.}
\end{figure}
Temperature-dependent magnetic susceptibility $\chi~(\equiv M/H$) of the polycrystalline Na$_3$Fe(PO$_4$)$_2$ sample measured in 0.5~T applied field is shown in Fig.~\ref{Fig4}(a). The most significant feature in $\chi(T)$ is the presence of a broad maximum at $T^{\rm max}_\chi \simeq 15$~K, mimicking the AFM short-range order, typical for low-dimensional spin systems. While going further below this broad maxima, a change in slope is observed at $T_{\rm N} \simeq 10.4$~K, which marks an alteration towards an AFM long-range-order (LRO). The ordering temperature is very well evident in the $d\chi/dT$ vs $T$ plot (not shown). To understand the nature of the magnetic ordering, $\chi(T)$ was measured at different applied fields upto 7~T [see the inset of Fig.~\ref{Fig4}(b)]. With increasing $H$, $T_{\rm N}$ remains unchanged upto 3~T and then moves weakly towards high temperatures above 3~T. For $H \geq 3$~T, $\chi(T)$ below $T_{\rm N}$ develops into a contort, which emphasizes that there is some field-induced spin canting prevailing in the system.

The zero-field cooled (ZFC) and field-cooled (ZFC) susceptibilities as a function of temperature were measured in a low field of 0.01~T (not shown here). The absence of splitting between ZFC and FC $\chi(T)$s certainly rules out the possibility of any spin-glass transition or spin freezing at low temperatures.

For the purpose of analysis, $\chi(T)$ data in the high temperature regime were fitted using the Curie-Weiss (CW) law,
\begin{equation}
\chi(T) = \chi_0 + \frac {C}{T+\theta_{\rm CW}},
	\label{eq3}
\end{equation}
where $\chi_{0}$ is the temperature-independent susceptibility, $C$ is the Curie constant, and $\theta_{\rm CW}$ is the characteristic CW temperature. The fit shown in Fig.~\ref{Fig4}(b) for $T \geq 150$~K yields the parameters: $\chi_{0} \simeq -5.25 \times 10^{-4}$~cm$^{3}$/mol, $C \simeq 4.9906$~cm$^{3}$K/mol, and $\theta_{\rm CW} \simeq 37.01$~K. A positive value of $\theta_{\rm CW}$ indicates the presence of dominate AFM interactions between the Fe$^{3+}$ ions. From the $\theta_{\rm CW}$ and $T_{\rm N}$ values, we can quantify frustration by the frustration ratio, $f$~($=\left| \theta_{\rm CW}/T_{N} \right|$)~$\simeq 3.6$, which apprises us that the system is moderately frustrated. The core diamagnetic susceptibility $\chi_{\rm core}$ of Na$_3$Fe(PO$_4)_2$ was calculated to be $-1.23 \times 10^{-4}$~cm$^{3}$/mol by summing the core diamagnetic susceptibilities of individual ions Na$^{+}$, Fe$^{3+}$, P$^{5+}$, and O$^{2-}$~\cite{Selwood2013,*Mendelsohn1130}. The Van-Vleck paramagnetic susceptibility ($\chi_{\rm VV}$), which arises from the second-order contribution to free energy in the presence of magnetic field was obtained by subtracting $\chi_{\rm core}$ from $\chi_0$ to be $\sim -4.03 \times 10^{-4}$~cm$^{3}$/mol.
Further, from the value of Curie constant $C$, the effective moment calculated using the relation $\mu_{\rm eff} = \sqrt{3k_{\rm B}C/N_{\rm A}}$ to be $\mu_{\rm eff} \simeq 6.319 \mu_{\rm B}$, where $k_{\rm B}$ is the Boltzmann constant, $\mu_{\rm B}$ is the Bohr magneton, and $N_{\rm A}$ is the Avogadro's number. For a spin-$5/2$ system, the spin-only effective moment is expected to be $\mu_{\rm eff} = g\sqrt{S(S+1)}\mu_{\rm B} \simeq 5.916$~$\mu_{\rm B}$, assuming Land\'e $g$-factor $g = 2$. However, our experimental value of $\mu_{\rm eff} \simeq 6.31$~$\mu_{\rm B}$ is slightly higher than the spin-only value and corresponds to a $g$-factor of $g \simeq 2.15$, which is consistent with our spin model fit discussed later. In the mean-field approximation, $\theta_{\rm CW}$ is a measure of energy scale of the total exchange interactions which is related to the exchange coupling as $|\theta_{\rm CW}|=\frac{JzS(S+1)}{3k_{\rm B}}$, where $z=6$ is the number of nearest-neighbours of Fe$^{3+}$ ions and $J$ is the average intra-layer exchange coupling~\cite{Domb296}. In this way, we estimated the average exchange coupling within the triangular planes to be $J/k_{\rm B} \simeq 2$~K.

Further, for a direct estimation of exchange coupling between the Fe$^{3+}$ ions and to understand the spin-lattice, we decomposed $\chi(T)$ into two components,
\begin{equation}
\chi(T)=\chi_0 + \chi_{\rm spin}(T).
\label{eq4}
\end{equation}
Here, $\chi_{\rm spin}(T)$ typifies the intrinsic spin susceptibility which can be chosen according to the intrinsic magnetic model. As the Fe$^{3+}$ ions in the crystal structure are arranged in triangles, we took the expression of high temperature series expansion (HTSE) of $\chi_{\rm spin}$ for a spin-$5/2$ Heisenberg isotropic TLA model which has the form\cite{Delmas55}
\begin{equation}
\frac{N_{A}\mu_{B}^{2}g^{2}}{3 \left| J\right|  \chi_{\rm spin}}= x + 4+\frac{3.20}{x}-\frac{2.186}{x^2}+\frac{0.08}{x}+\frac{3.45}{x^4}-\frac{3.99}{x^5},
\label{eq5}
\end{equation}
with $x= k_{\rm B}T/\lvert J \rvert S(S+1)$. This expression is valid for $T \geq JS(S+1)$~\cite{Schmidt104443}. The solid line in Fig.~\ref{Fig4}(a) represents the best fit to the $\chi(T)$ data above 20~K by Eq.~\eqref{eq4} resulting $\chi_{0} \simeq -6.22 \times 10^{-4}$~cm$^{3}$/mol, $g \simeq 2.11$, and the average AFM exchange coupling $J/k_{\rm B} \simeq 1.8$~K. Indeed, this value of $J/k_{\rm B}$ is in good agreement with the value estimated from $\theta_{\rm CW}$.
We also tried to fit the $\chi(T)$ data using HTSE for $\chi_{\rm spin}(T)$ of the spin-$5/2$ Heisenberg 1D chain model in Eq.~\eqref{eq4}~\cite{Dingle643}. The fit however did not converge giving unphysical parameters. Thus, the agreement of our experimental $\chi(T)$ data with the isotropic TLA model suggests frustrated triangular spin-lattice in the system.

\begin{figure}
\includegraphics[width=\columnwidth] {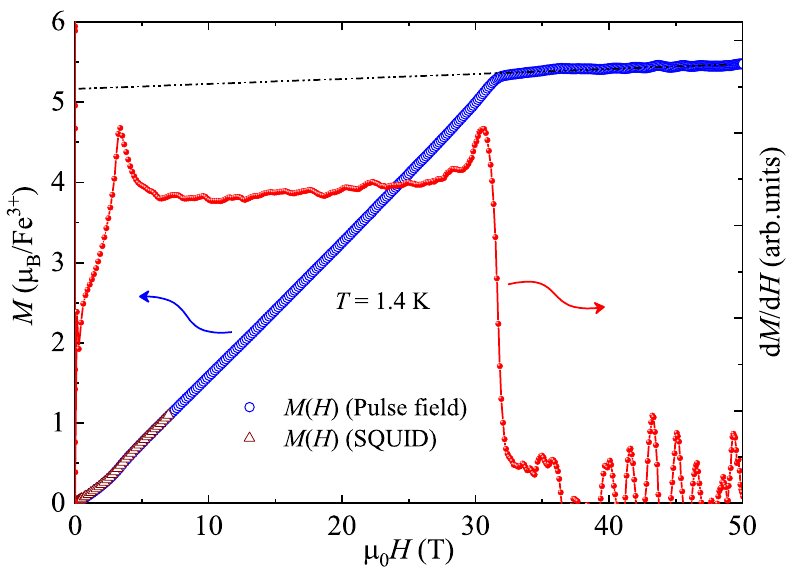}
\caption{\label{Fig6} Magnetization ($M$) and its field derivative $dM/dH$ vs $H$ measured at $T = 1.4$~K using pulsed magnetic field in the left and right $y$-axes, respectively. The pulse field data are scaled with respect to the SQUID data measured up to 7~T. The dash-dotted line represents the saturation magnetization $M_{\rm sat} \sim 5.2$~$\mu_{\rm B}$ per Fe$^{3+}$ ion. The sharp peaks in $dM/dH$ vs $H$ plot depict the spin-flop and saturation fields $\mu_0 H_{\rm SF}\simeq3.2$~T and $\mu_0 H_{\rm sat}\simeq31.5$~T, respectively.}
\end{figure}
To look for the field-induced transition, we measured the magnetization isotherms ($M$ vs $H$) at different temperatures up to 7~T. As shown in Fig.~\ref{Fig5}(a), for $T < 10$~K a bend is observed in the intermediate magnetic fields which is the signature of a metastable field-induced transition. It is more pronounced in the $dM/dH$ vs $H$ plots shown in Fig.~\ref{Fig5}(b) where this feature is manifested in a well defined peak. As the temperature rises, one can observe that the peak is moving slightly towards higher fields and then disappears completely for $T > 9$~K.

Further details about field induced transition and saturation magnetization can be obtained from the high field magnetization measurements. The $M$ vs $H$ data measured in pulsed magnetic field up to 50~T at $T = 1.4$~K are plotted in Fig.~\ref{Fig6}. The pulse field magnetization curve is quantified by scaling it with respect to the SQUID data measured up to 7~T at $T = 1.8$~K. It exhibits a kink at $\mu_0 H_{\rm SF}\simeq3.2$~T, similar to that observed in Fig.~\ref{Fig5}, reminiscent of a spin-flop (SF) transition. Above $H_{\rm SF}$, $M$ increases almost linearly with $H$ and develops a sharp bend towards saturation above $\mu_0 H_{\rm sat} \simeq 31.5$~T. These two critical fields are very well visualized as sharp peaks in the $dM/dH$ vs $H$ plot. No obvious features associated with the $1/3$ magnetization plateau is observed in the intermediate field range. In the polycrystalline sample, the random orientation of the grains with respect to the applied field direction might
obscure this plateau. The linear interpolation of a straight line fit to the magnetization above $H_{\rm sat}$ intercepts $y$-axis at $M_{\rm sat}\sim5.25$~$\mu_{\rm B}$. This saturation magnetization corresponds to the value expected for a spin-$5/2$ (Fe$^{3+}$) ion with $g= 2.1$. Further, in a spin system $H_{\rm sat}$ defines the energy required to overcome the AFM interactions and to polarize the spins in the direction of applied field. In particular, in a Heisenberg TLA, $H_{\rm sat}$ can be written in terms of the intra-layer exchange coupling as $\mu_0 H_{\rm sat} = 9JS/g\mu_{\rm B}$~\cite{Kawamura4530}. Our experimental value of $\mu_0 H_{\rm sat} \simeq 31.5$~T yields an average exchange coupling of $J/k_{\rm B} \simeq 1.95$~K which is indeed close to the value obtained from the analysis of $\chi(T)$ and $\theta_{\rm CW}$. A small difference can be attributed to the magnetic anisotropy present in the compound.

\subsection{Heat Capacity}
\begin{figure}
	\centering
	\includegraphics[width=\columnwidth]{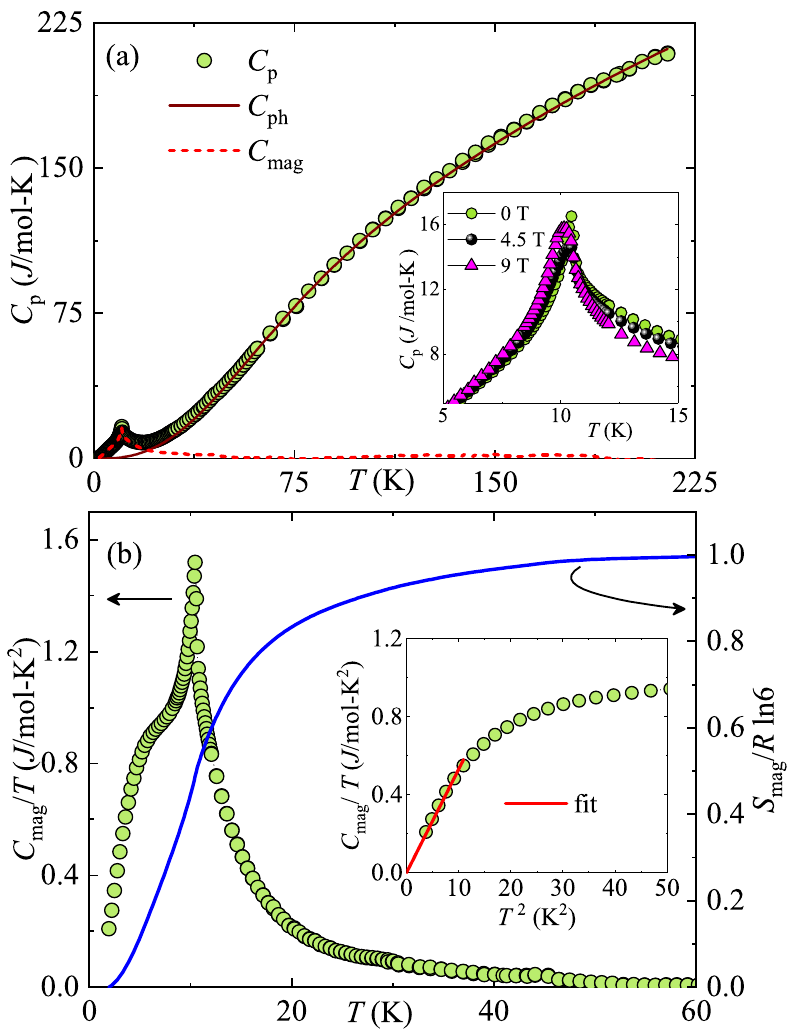}
	\caption{\label{Fig7} (a) Heat capacity ($C_{\rm p}$) as a function of temperature in zero applied field. The solid line represents the phonon contribution ($C_{\rm ph}$) [Eq.~(6)] while the dashed line indicates the magnetic contribution ($C_{\rm mag}$). Inset: Field dependence of $C_{\rm p}(T)$ around $T_{\rm N}$. (b) $C_{\rm mag}/T$ and $S_{\rm mag}$ vs $T$ in the left and right $y$-axes, respectively. Inset: $C_{\rm mag}/T$ vs $T^{2}$ and the solid line is the linear fit.}
\end{figure}
The temperature-dependent heat capacity $C_{\rm p}$ of the polycrystalline sample measured in zero field is shown in Fig.~\ref{Fig7}(a). As we go lower in temperature, $C_{\rm p}$ also goes down systematically and then shows a pronounced $\lambda$-type anomaly at $T_{\rm N}\simeq10.4$~K, illustrating the crossover to a  magnetically ordered state. Typically, in high temperatures, $C_{\rm p}(T)$ in magnetic insulators has a dominant contribution from phonon excitations ($C_{\rm ph}$), whereas in low-temperatures, the magnetic part of the heat capacity ($C_{\rm mag}$) predominates over $C_{\rm ph}$. Therefore, in magnetic systems with low energy scale of the exchange coupling, one can separate the magnetic part from the phononic part by analyzing $C_{\rm p}$ in the high temperature regime.

Inorder to demarcate $C_{\rm mag}$ from the total heat capacity $C_{\rm p}$, the phonon contribution was first estimated by fitting the high-$T$ data by a linear combination of one Debye and three Einstein terms [Debye-Einstein (DE) model] as~\cite{Gopal2012,Sebastian064413}
%which is further subtracted from $C_{\rm p} (T)$. Mathematically, one can express the Debye-Einstein model as~\cite{Gopal2012,Sebastian064413} 
\begin{equation}
	C_{\rm ph}(T)=f_{\rm D}C_{\rm D}(\theta_{\rm D},T)+\sum_{i = 1}^{3}g_{i}C_{{\rm E}_i}(\theta_{{\rm E}_i},T).
	\label{Eq6}
\end{equation}
The first term in Eq.~\eqref{Eq6} represents the Debye model,
\begin{equation}
	C_{\rm D} (\theta_{\rm D}, T)=9nR\left(\frac{T}{\theta_{\rm D}}\right)^{3} \int_0^{\frac{\theta_{\rm D}}{T}}\frac{x^4e^x}{(e^x-1)^2} dx,
	\label{Eq7}
\end{equation}
where, $x=\frac{\hbar\omega}{k_{\rm B}T}$, $\omega$ is the frequency of oscillation, $R$ is the universal gas constant, and $\theta_{\rm D}$ is the characteristic Debye temperature. The flat optical modes in the phonon spectra is accounted by the second term in Eq.~\eqref{Eq6}, known as the Einstein term
\begin{equation}
	C_{\rm E}(\theta_{\rm E}, T) = 3nR\left(\frac{\theta_{\rm E}}{T}\right)^2 
	\frac{e^{\left(\frac{\theta_{\rm E}}{T}\right)}}{[e^{\left(\frac{\theta_{\rm E}}{T}\right)}-1]^{2}},
	\label{Eq7} 
\end{equation}
where, $\theta_{\rm E}$ is the characteristic Einstein temperature. In Eq.~\eqref{Eq6} one can tentatively assign $\theta_{\rm D}$ to the low-energy vibrations of the heavy atom (Fe) and the remaining three lighter atoms (Na, P, O) can be accounted separately by the three Einstein terms, respectively. The coefficients $f_{\rm D}$, $g_1$, $g_2$, and $g_3$ are the weight factors, which take into account the number of atoms per formula unit ($n$) and are chosen in such a way that at high temperatures the Dulong-Petit value ($\sim 3nR$) is satisfied~\cite{Fitzgerel545}.

The zero field $C_{\rm p}(T)$ data above $\sim 20$~K are fitted by Eq.~\eqref{Eq6} [see, solid curve in Fig.~\ref{Fig7}(a)] and the obtained parameters are $f_{\rm D} \simeq 0.075$, $g_1 \simeq 0.205$, $g_2 \simeq 0.30$, $g_3 \simeq 0.42$, $\theta_{\rm D} \simeq 155$~K, $\theta_{{\rm E}_1} \simeq 220$~K, $\theta_{{\rm E}_2} \simeq 360$~K, and $\theta_{{\rm E}_3} \simeq 930$~K. One may notice that the sum of $f_{\rm D}$, $g_1$, $g_2$, and $g_3$ is close to one, as expected. Finally, the high-$T$ fit was extrapolated down to 2~K and subtracted from $C_{\rm p}(T)$ to get $C_{\rm mag}(T)$ [see, Fig.\ref{Fig7}(a)]. Figure~\ref{Fig7}(b) presents $C_{\rm mag} (T)/T$ and the corresponding magnetic entropy [$S_{\rm{mag}}(T) = \int_{\rm 2\,K}^{T}\frac{C_{\rm {mag}}(T')}{T'}dT'$] vs $T$. The obtained magnetic entropy which saturates above 50~K approaches a value $S_{\rm{mag}}\simeq 14.85$~J/mol-K which is close to the expected theoretical value $S_{\rm{mag}}= R ln(2S+1)= 14.89$~J/mol K for a $S= 5/2$ system.

$C_{\rm P}(T)$ measured in different applied fields in the low temperature regime is shown in the inset of Fig.~\ref{Fig7}(a). No noticeable shift in $T_{\rm N}$ is apparent for a field change from 0 to 9~T. Further, well below $T_{\rm N}$, $C_{\rm mag}(T)$ follows a $T^3$ behaviour [inset of Fig.~\ref{Fig7}(b)], depicting the dominance of three-dimensional (3D) magnon excitations~\cite{Bag134410,Somesh104422}. 
 
\subsection{Neutron diffraction}
\begin{figure}
	\centering
	\includegraphics[width=\columnwidth]{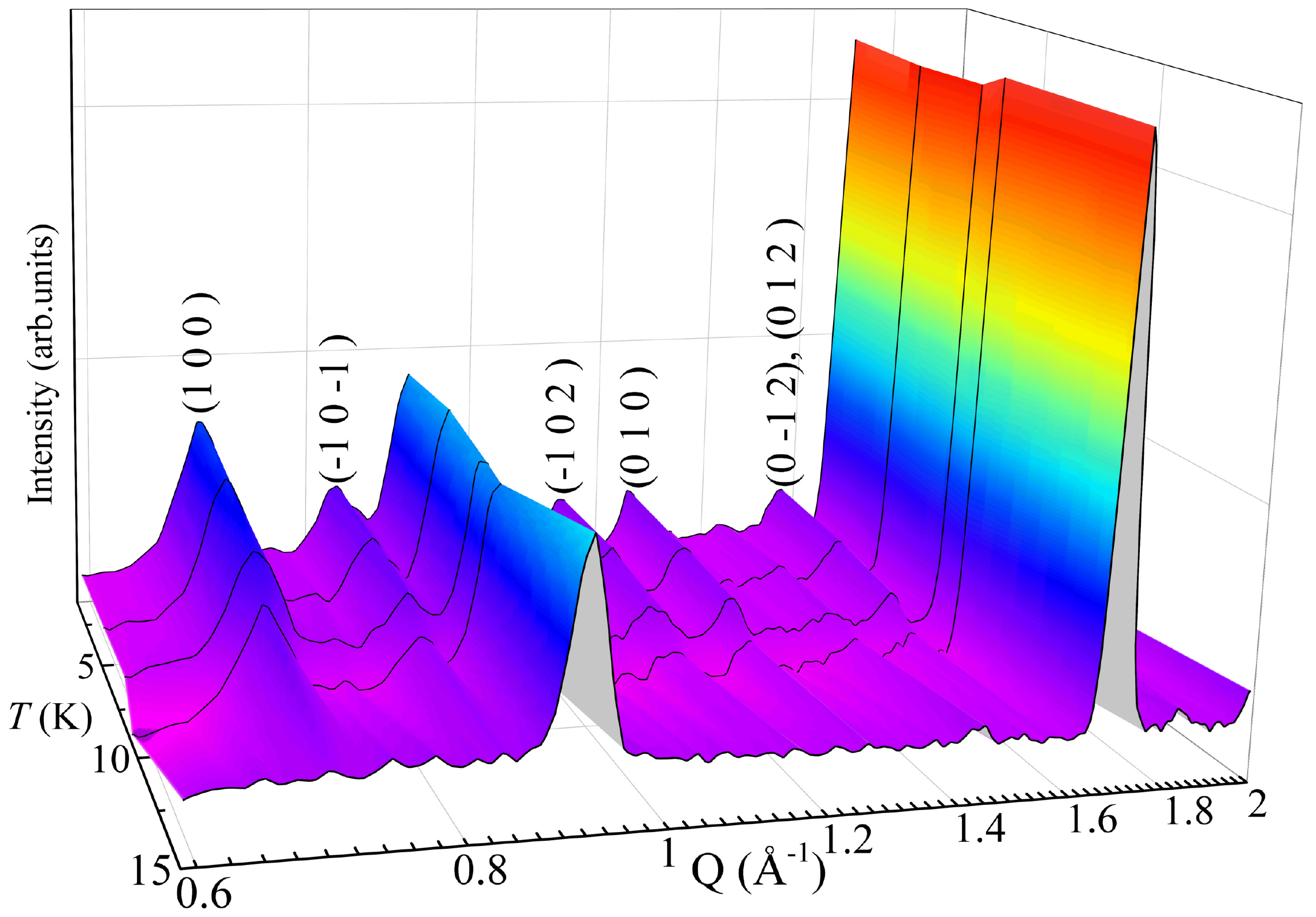}
	\caption{\label{Fig8} (a) Temperature evolution of the neutron powder diffraction (NPD) patterns of Na$_3$Fe(PO$_4$)$_2$ in the low-$Q$ regime and for temperatures below 15~K. An emergence of magnetic reflections with $k$ = $(1, 0 ,0)$ is evident below $T_{\rm N}$.}
\end{figure}
\begin{figure}
\centering
\includegraphics[width=\columnwidth]{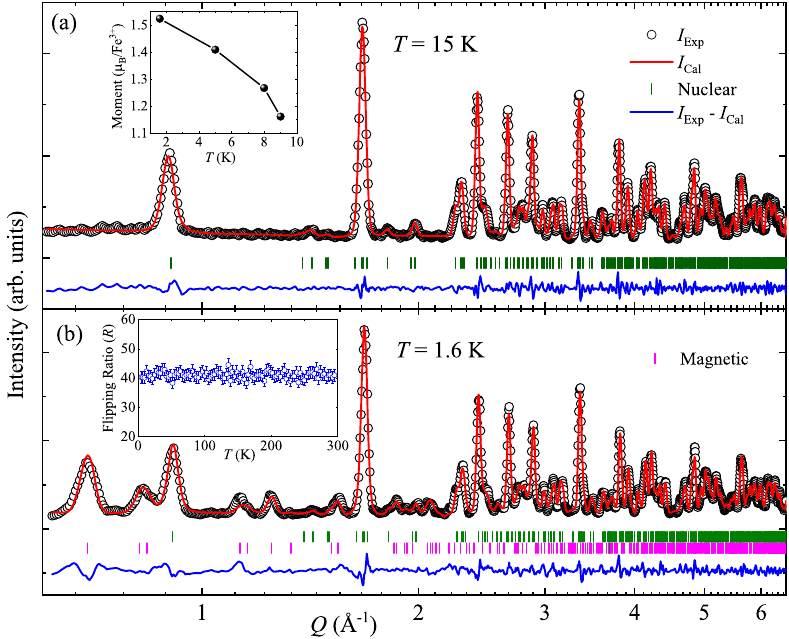}
\caption{\label{Fig9} Rietveld refinement of the neutron powder diffraction data at (a) $T=15$~K and (b) $T=1.6$~K, considering only nuclear reflections and nuclear~+~magnetic reflections, respectively. The experimental data are depicted as open black circles. The red solid line represents the calculated pattern. The difference in intensities of experimental and calculated data is shown as blue solid line at the bottom. The allowed Bragg peaks corresponding to the nuclear (green) and magnetic (pink) reflections are shown as vertical bars. Inset of (a) represents the temperature variation of ordered magnetic moment, below $T_{\rm N}$. Inset of (b) represents the variation of flipping ratio ($R$) in the $T$-range 5~K$\leq T\leq$300~K.} 
\end{figure}

The temperature evolution of the NPD pattern at different temperatures (1.6 to 15~K) are shown in Fig.~\ref{Fig8}. New peaks were found to appear below $T_{\rm N}$ at $Q \sim $~0.6921, 0.8180, 1.1227, 1.2503, and 1.5455~\AA$^{-1}$. As we reduce the temperature, the intensity of these peaks increases gradually, suggesting the emergence of magnetic reflections. The appearance of new magnetic reflections suggests the AFM nature of the transition~\cite{Anand094402,*Islam134433}. In Fig.~\ref{Fig8}, the $x$-axis is shown in log-scale in order to highlight these magnetic reflections in the low-$Q$ (momentum transfer vector) regime. The magnetic peaks developed below $T_{\rm N}$ can be indexed with an ordering vector $k = (1, 0, 0)$, which insinuates a commensurate magnetic ordering. The Rietveld refinement of the nuclear pattern at $T=15$~K was performed considering monoclinic crystal structure (space group: $C2/c$). The best fit of the data [see, Fig.~\ref{Fig9}~(a)] yields $a = 9.1085(2)$~{\AA}, $b = 5.0532(1)$~{\AA}, $c = 13.8733(3)$~{\AA}, $\beta = 91.32(1)^{\circ}$, and $V_{\rm cell}\simeq 638.4$~{\AA}$^{3}$. These obtained lattice parameters are in close agreement with the refined values from the powder XRD data at room temperature~\cite{Morozov377}.

In the crystal structure of Na$_3$Fe(PO$_4$)$_2$, magnetic Fe-atom occupies $4a$ $(0,0,01)$ crystallographic site. For determining the magnetic structure the standard representational theory was used~\cite{Bertaut149}. For the propagation vector $k = (1, 0, 0)$, the little group $G_{k}$ can be generated using two symmetry elements, $-x,y,-z+1/2$ and $-x,-y,-z$. Our symmetry analysis shows that the magnetic representation for $4a$ site can be written in terms of the irreducible representations (IR) as $\Gamma (4a)=3\Gamma _{1}^{(1)}+3\Gamma _{3}^{(1)}$. Here, both IRs ($\Gamma _{1}$ and $\Gamma _{3}$) are one-dimensional and are composed of basis vectors $\Psi_{1, 2, 3}$ and $\Psi_{4, 5, 6}$, respectively (see, Table~\ref{table1}).
%Additional peaks appear in the neutron diffraction pattern at momentum transfer vector ($Q$) value of 0.6921, 0.8180, 1.1227, 1.1586, 1.2503, and 1.5455~\AA$^{-1}$ for $T\leq9$~K, indicating the presence of long-range antiferromagnetic ordering, in agreement with specific heat and magnetization measurements.
%The magnetic structure has been determined using the standard representational theory~\cite{Bertaut149}. For the propagation vector $k$ = $(1, 0, 0)$, little group $G_{k}$ can be generated using the 2 elements, $-x,y,-z+1/2$ and $-x,-y,-z$.Symmetry analysis shows that the magnetic representation for site $4a$ can be written as $\Gamma (4a)=3\Gamma _{1}^{(1)}+3\Gamma _{3}^{(1)}$. 

\begin{figure}
\centering
\includegraphics[width=\columnwidth]{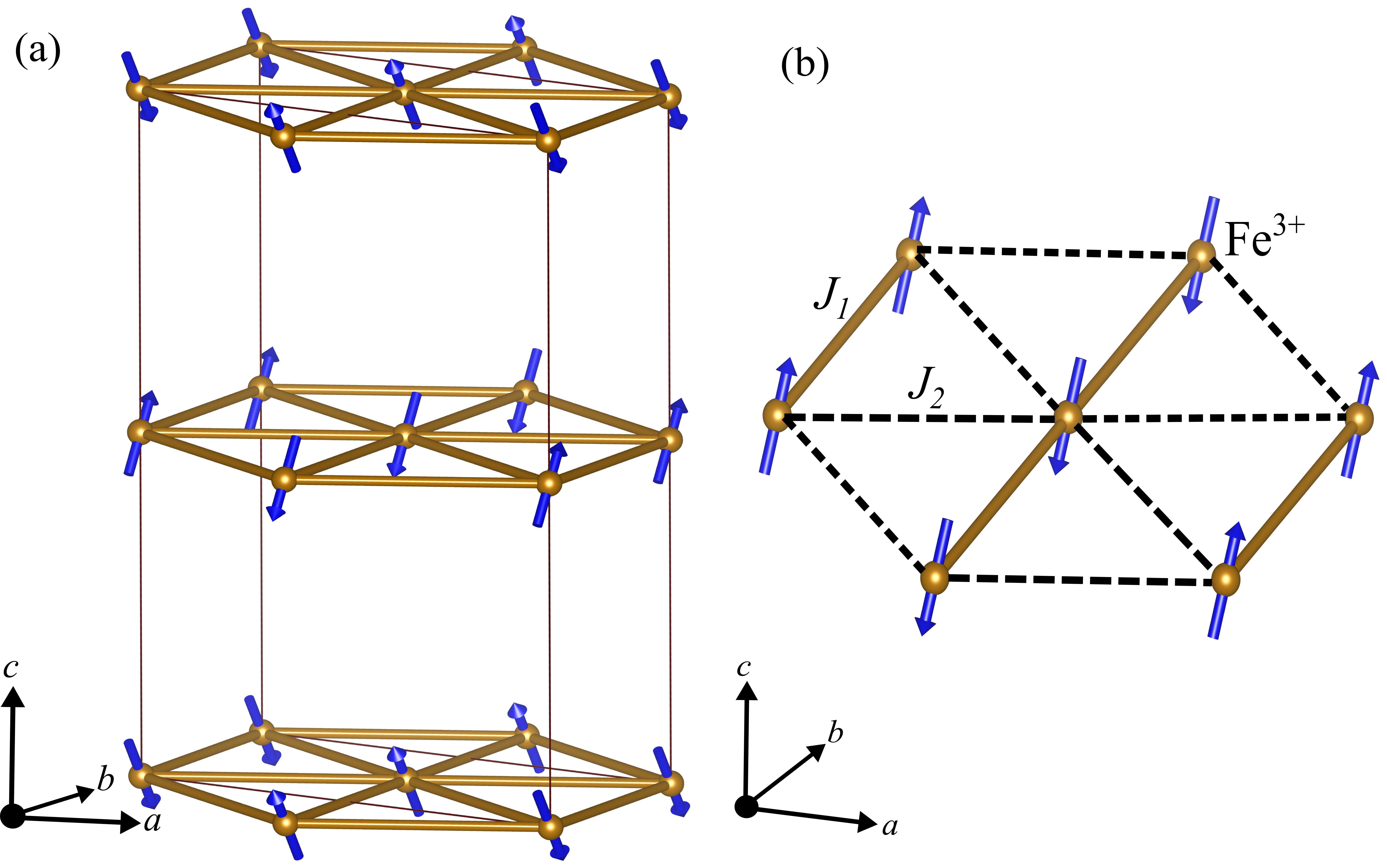}
\caption{\label{Fig10} (a) Spin structure of Na$_3$Fe(PO$_4$)$_2$ derived from the neutron powder diffraction at $T= 1.6$~K. (b) A section of the triangular layer ($ab$-plane) showing the spin orientation and exchange interactions.}
\end{figure}

The observed NPD patterns cannot be fitted with the magnetic structure corresponding to the representations $\Gamma _{1}$ . Whereas, the magnetic structure corresponding to the representation $\Gamma _{3}$ is found to be appropreate to describe the observed NPD patterns. The magnetic structure belonging to the representation $\Gamma _{3}$ corresponds to an antiferromagnetic structure, where moments are aligned in the $bc$-plane with dominant component along the crystallographic $c$-axis (Fig.~\ref{Fig10}). Further details of the spin structure is discussed in the next section. From the magnetic structure refinement, we were able to quantify the magnetic moment values along different crystalographic directions. At the lowest temperature ($T=1.6$~K), these values are $\mu_{\rm b} \simeq 0.83~\mu_{\rm B}$ and $\mu_{\rm c} \simeq 1.27~\mu_{\rm B}$ along the $b$- and $c$-directions, respectively. The temperature variation of total moment ($\mu = \sqrt{\mu_b^2 + \mu_c^2}$) is shown in the inset of Fig.~\ref{Fig9}(a). The observed reduced value of ordered moment with respect to the theoretically expected classical value of $5\mu_{B}$ for spin-$5/2$ could be attributed to the effect of low-dimensionality and magnetic frustration and/or presence of covalence effect~\cite{Sanjeewa2813}.

Further, to check the presence of any ferromagnetic-type correlations, we have also carried out the neutron depolarization measurements. The absence of depolarization is apparent from the flipping ratio ($R$) vs $T$ plot, where $R$ is found to be constant over the entire temperature range (5~K~$\leq T \leq 300$~K) [see, inset of Fig.~\ref{Fig9}(b)]. This rules out the possibility of any ferromagnetic-type correlations or spin canting in the system.

%Further, to know the presence of any ferromagnetic type correlations, we also have carried out the neutron depolarization measurements. No depolarization of the neutron was noticed, confirming the absence of ferromagnetic-type correlations.
%The variation of the refined value of the moment at various temperature are shown in inset of (a) Fig~\ref{Fig8}. At $T=1.6$~K, the magnetic moments lie in the $ab$ plane with $\mu_{\rm b}=0.834~\mu_{B}$ and $\mu_{\rm b}=1.275~\mu_{B}$ respectively.
%This reduced value of the moment compared to the theoretically expected value of 5$\mu_{B}$ could be due to geometrical frustration or covalence effects. Further, to know the presence of any ferromagnetic type correlations, we also have carried out the neutron depolarization measurements. No depolarization of the neutron was noticed, confirming the absence of ferromagnetic-type correlations.
\begin{table}
\caption{\label{table1} Irreducible representations and basis vectors for the space group $C2/c$ with propagation vector $k$ = $(1, 0, 0)$, obtained using the BasIreps software. The atoms are defined according to Fe1: (0, 0, 0) and Fe2: (0, 0, 0.5).}
\begin{ruledtabular}
\begin{tabular}{c@{\hspace{-1.5em}}cccc}
\multicolumn{5}{c}{Basis vector components for $k$ = $(1, 0, 0)$} \\
\hline
\vspace{2pt}
& IR & Basis Vectors & Fe1 & Fe2 \\
\hline 
& $\Gamma _{1}$    &  $\Psi_{1}$  &   (1 0 0)   & (-1 0 0) \\
&                  &  $\Psi_{2}$  &   (0 1 0)   & (0 1 0) \\
&                  &  $\Psi_{3}$  &   (0 0 1)   & (0 0 -1) \\
\hline
& $\Gamma _{3}$    &  $\Psi_{4}$  &   (1 0 0)   & (1 0 0) \\
&                  &  $\Psi_{5}$  &   (0 1 0)   & (0 -1 0) \\
&                  &  $\Psi_{6}$  &   (0 0 1)   & (0 0 1) \\
\end{tabular}
\end{ruledtabular}
\end{table}

\section{Discussion and Summary}
\label{Discussion}
Theoretically, for a spin-$5/2$ chain system, the broad maxima in $\chi(T)$ is expected at $k_{\rm B} T_{\chi}^{\rm max}/J =10.6$~\cite{Jongh947}. The broad maximum in our experimental data appears at $T_{\chi}^{\rm max} \simeq 15$~K which corresponds to the intra-chain coupling of $J/k_{\rm B}\simeq 1.41$~K. In order to reproduce the overall shape of the $\chi(T)$ plot, we simulated the data using QMC technique, assuming a spin-chain model with $S=5/2$ and $J/k_{\rm B}\simeq 1.41$~K. As depicted in the inset of Fig.~\ref{Fig4}(a), the simulated data show a significant deviation from the experimental data, discarding the possibility of spin-chain physics. On the other hand, the fit assuming isotropic triangular lattice [Eq.~(\ref{eq4})] reproduces the experimental $\chi(T)$ data very well in the high temperature regime, despite a small spatial anisotropy. As a further test for the isotropic TLA model, we performed the full diagonalization for a spin-$5/2$ isotropic TLA with $J/k_{\rm B}\simeq 1.8$~K [obtained from the $\chi(T)$ analysis]. As evident from the inset of Fig.~\ref{Fig4}(a), the simulated data using full diagonalization superimpose the experimental data, above $T_{\chi}^{\rm max}$. This confirms that the spin-lattice in Na$_3$Fe(PO$_4$)$_2$ behaves more of a 2D triangular lattice, rather than a 1D spin-chain.

This compound orders antiferromagnetically below $T_{\rm N}\simeq10.4$~K which is clear from the appearance of magnetic reflections in the NPD pattern, $\lambda$-type anomaly in $C_{\rm p}(T)$, and slope change in $\chi(T)$. The magnetic structure refined from the NPD data is shown in Fig.~\ref{Fig10} which demonstrates a collinear AFM ordering. The spins are coupled ferromagnetically ($J_1$) along the $b$-axis or in the [100] plane which is the direction of the shortest Fe$^{3+}$-Fe$^{3+}$ distance. The AFM interchain coupling ($J_2$) ties the chains into a 2D anisotropic triangular layer in the $ab$-plane. Alternatively, one can visualize it as AFM cross chains running along the [110] and [-110] directions which are coupled ferromagetically in the $ab$-plane.
%Further, it stabilizes in a collinear AFM ordering instead of a non-collinear 120$^{\circ}$ order expected for an isotropic TLA.
Though a non-collinear 120$^{\circ}$ structure is understood as a hallmark of TLA, some compounds tend to deviate from this spin structure either due to anisotropy or inter-layer coupling~\cite{Ranjith094426,*Ranjith024422}.

The crystal symmetry plays a key role in deciding the ground states of the compounds in the $AA^{\prime}M$($X$O$_4$)$_2$ family.
%The compounds, BaNa$_2$Co[(P,V)O$_4$]$_2$~\cite{Wellm100420,Sanjeewa2813}, (K, Rb)Ag$_2$Cr(VO$_4$)$_2$~\cite{Tapp064404,Lee224420}, and (K,Rb)Ag$_2$Fe(VO$_4$)$_2$~\cite{Amuneke5930} which crystallize in a higher symmetry space groups $P\bar{3}m1$, $P\bar 3$, and $P\bar 3$, respectively are having equal bond distances leading to isotropic ($J'/J=1$) triangular lattices.
%In these compounds, the ratio of intra-chain to inter-chain exchange interactions is unity (i.e. $J'/J=1$), placing themself at the extreme horizon of the $J - J'$ phase diagram~\cite{Weichselbaum245130}.
%Indeed, the 120$^{\circ}$ noncollinear ordering is perceived for (K,Rb)Ag$_2$Fe(VO$_4$)$_2$ (space group: $P\bar 3$) from the neutron diffraction experiments~\cite{Amuneke5930}.
The correspondence between crystal symmetry and exchange couplings is well demonstrated for the series $A$Ag$_2$Cr(VO$_4$)$_2$ ($A = $~Ag, K, Rb)~\cite{Tapp064404}. Here, replacement of $A$ site induces a symmetry change in the CrO$_6$ octahedra. Therefore, Ag$_3$Cr(VO$_4$)$_2$ (space group: $C2/c$) is a distorted/anisotropic TLA with a collinear AFM ordering below $T_{\rm N} \simeq 10$~K, whereas the other compounds ($A = $~K, Rb) with higher symmetry (space group: $P\bar 3$) are undistorted/isotropic TLAs and are reported show signature of QSL~\cite{Tapp064404}. Several other compounds with higher symmetry in this series also feature isotropic triangular lattices~\cite{Wellm100420,Sanjeewa2813,Tapp064404,Lee224420,Amuneke5930}.
%For compound with even lower symmetry space group ($C2/c$), like in BaNa$_2$Mn(VO$_4$)$_2$, which is recognized as a TLA with slight distortion, the spins are found to order antiferromagnetically~\cite{Sanjeewa2813}.
%BaAg$_2$Cu(VO$_4$)$_2$, a sister compound of the same family crystallizes in a triclinic space group, $P\bar 1$ (low symmetry space group) was initially proposed to be a spin-$1/2$ anisotropic triangular lattice system, due to its anisotropic Cu--Cu bond lengths~\cite{Amuneke2207}. Subsequently, the band structure calculations ruled out this proposal and established a superposition of ferromagnetic and antiferromagnetic uniform spin chains, placing itself more close to the a pure spin chain regime ($J'/J= 0$) in the phase diagram~\cite{Tsirlin014401}.
%Very recently, a Cu-based spin-$1/2$ compound BaNa$_2$Cu(VO$_4$)$_2$ [monoclinic space group ($C2/c$)] has gone through some significant experimental and theoretical analysis culminating the structure to be a spin-chain rather than an TLA~\cite{Sebastian064413}.
%Eventhough a considerable anisotropy in $M-M$ ions in bond distances, an iron-based (Fe$^{3+}$) TL compound, Ag$_3 $Fe(VO$_4$)$_2$ is classified as a distorted TL has provided insights on how the chemically induced octahedral distortions affects the thermodynamic as well as magnetic properties apart from the pure TL counterparts~\cite{Amuneke5930}.
As pointed out in Fig.~\ref{Fig1}(b), for Na$_3$Fe(PO$_4$)$_2$, the PO$_4$ tetrahedra which acts as a medium of interaction among the Fe$^{3+}$ ions is distorted where the P-O bond distance varies from 1.50~\AA to 1.56~\AA and the $\angle O-P-O$ angles are also different. These give rise to non-equivalent  exchange interactions along the edges of the triangle. The anisotropic interactions can also be assessed from the deviation of O-Fe-O bond angles from 90$^{\circ}$, typically expected for an isotropic triangular lattice formed by undistorted FeO$_6$ octahedra. It is also to be noted that despite the shortest Fe$^{3+}$-Fe$^{3+}$ distance, the interaction along the [100] direction is FM. A close inspection of the bond angles reveals that $\angle O-P-O$ along the [100] direction has the smallest value of 97.13$^{\circ}$ compared to the other directions which possibly favours parallel alignment.
The collinear order in Na$_3$Fe(PO$_4$)$_2$ may thus be attributed to the distortion in FeO$_6$ octahedra and inter-layer coupling along the stacking $c$-direction, similar to Ag$_3$Cr(VO$_4$)$_2$.

\begin{figure}
	\centering
	\includegraphics[width=\columnwidth]{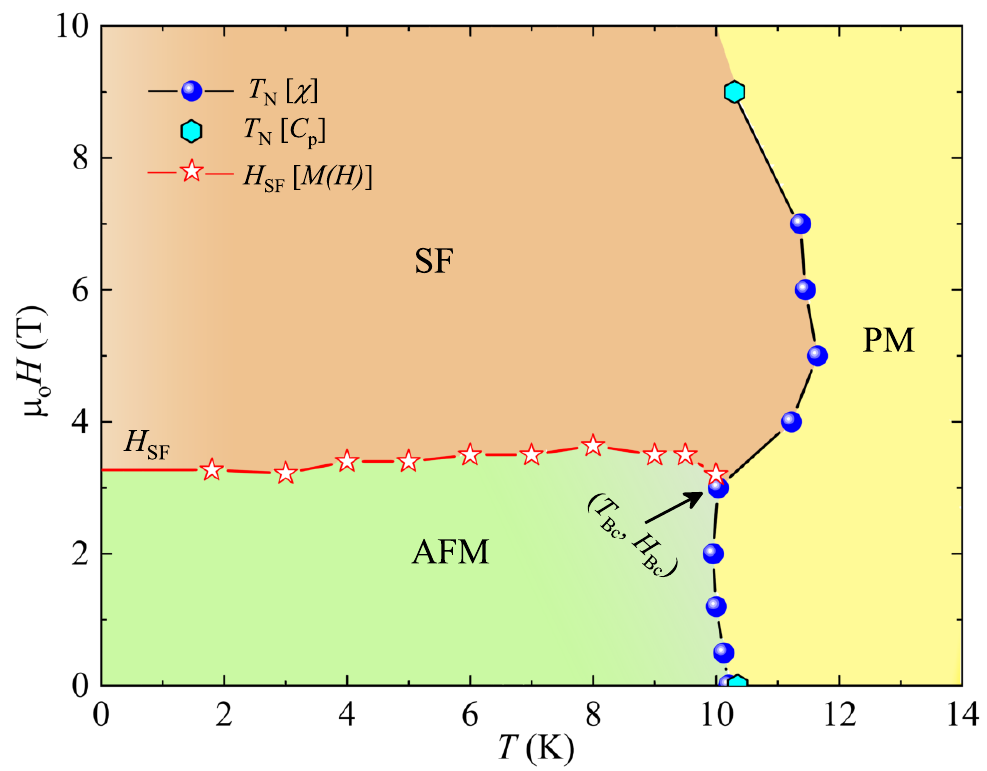}
	\caption{\label{Fig11} $H-T$ phase diagram of Na$_3$Fe(PO$_4$)$_2$ drawn from the magnetic isotherm, $\chi(T)$, and $C_{\rm p}(T)$ measurements. $H_{\rm SF}$ represents the critical field at which the SF transition takes place. The point ($T_{\rm Bc}, H_{\rm Bc})$ denotes the bi-critical point, where the SF transition ends.}
\end{figure}
The obtained $T_{\rm N}$ from $\chi(T)$ and $C_{\rm p}(T)$ data and the critical field ($H_{\rm SF}$) corresponding to the metastable transition obtained from the magnetic isotherms are summarized in Fig.~\ref{Fig11} as a $H-T$ phase diagram. It displays three well defined regions represented as paramagnetic (PM), antiferromagnetic (AFM), and spin-flop (SF) phases.
Majority of TLAs feature a magnetization plateau at $1/3$ of the saturation magnetization in the magnetic isotherms~\cite{Chubukov69,Ono104431,*Susuki267201,Smirnov134412,Hwang257205}, where the system evolves from a $120^{\degree}$ AFM order, which is the true ground state of a Heisenberg TLA~\cite{Capriotti3899} to a up-up-down ($uud$) phase. The $uud$ phase is stabilized by the magnetic field that introduces axial (Ising) anisotropy and disfavors the noncollinear $120^{\degree}$ order.
%Often, thermal fluctuations may also stabilize the collinear $uud$ phase in zero field resulting in two successive transitions in zero-field: $uud$ phase and the $120^{\degree}$ order, as in Ba$_3$(Mn,Co)Nb$_2$O$_9$~\cite{Lee224402,Lee104420}.
%Other TLAs reveal the uud phase in applied magnetic fields only~\cite{Smirnov134412,Hwang257205}.
%At higher fields, yet another, 0-coplanar or canted uud phase is stabilized~\cite{Seabra214418,Gvozdikova164209,Yamamoto027201}. The transition between the uud and 0-coplanar phases is typically weakly field dependent, so it can be paralleled to the II-III transition line.
Though our low temperature isotherms illustrate a plateau but it is well below the $1/3$ of the saturation magnetization~\cite{Seabra214418}. Further, the magnetic ordering in zero field is a collinear AFM order rather than a noncollinear $120^{\degree}$ order. Hence, the $1/3$ magnetization plateau is overruled. Moreover, the presence of a easy-plane (XY-type) anisotropy often triggers field emergent phenomenon like SF transition~\cite{Petrenko8983} above a critical field $H_{\rm SF}$, where the moments flip from parallel to perpendicular direction with respect to the applied field. Thus, the metastable/SF transition in Na$_3$Fe(PO$_4$)$_2$ can be attributed to the easy-plane anisotropy in the compound.
%where an increased magnetization in magnetic isotherm was realized resulted from flipping of moments parallel to perpendicular direction with respect to the applied field. The SF lines for these compounds were then accounted from a $H-T$ phase diagram obtained by measuring $M(H)$ at different temperatures.
Na$_2$BaMnV$_2$O$_8$, which belongs to the same series also shows a SF transition at low temperatures~\cite{Nakayama116003}. Similar phase diagram is also reported previously in the frustrated spin chain compounds SrCuTe$_2$O$_6$ and $\alpha$-Cu$_2$As$_2$O$_7$~\cite{Ahmed214413,*Arango134430}.

In summary, we have demonstrated the ground state properties of the frustrated magnet Na$_3$Fe(PO$_4$)$_2$. Based on the structural data one may anticipate a small anisotropy in the exchange couplings in the triangular unit. However, the analysis of our experimental data emphasizes that the system behaves more like a spin-$5/2$ isotropic triangular lattice. It stabilizes in a collinear antiferromagnetic ordering below $T_{\rm N}$ likely due to moderate inter-layer coupling. The occurrence of a field induced SF transition in the low temperature magnetic isotherms insinuates the presence of XY-type anisotropy. A comparison with other members of the $AA^{\prime}M$($X$O$_4$)$_2$ family reveals that the bond distances between metal ions and the corresponding angles become more anisotropic with reduced symmetry, leading to a potential structural complexity and exotic ground states.

\acknowledgements
SJS and RN would like to acknowledge SERB, India for financial support bearing sanction Grant No.~CRG/2019/000960. SJS is supported by the Prime Minister’s Research Fellowship (PMRF) scheme, Government of India. We also acknowledge the support of the HLD at HZDR, member of European Magnetic Field Laboratory (EMFL).
	
%\bibliography{ref_NFP}

	
\end{document}